# Transient Stability of Power Systems Integrated With Inverter-Based Generation

Xiuqiang He, *Student Member, IEEE*, Hua Geng, *Senior Member, IEEE*

***Abstract*—The fundamental mechanism of transient stability of the conventional power system can be physically understood with the rotor swing dynamics of synchronous generators (SGs). The basic factor affecting the transient stability is the power-angle relationship among the SGs. The integration of a large number of inverter-based generations (IBGs) into the power system inevitably changes the original power-angle relationship, therefore leading to a more complicated transient stability mechanism. In this letter, the impact of IBG on the power system transient stability is analytically investigated. It is shown that the connection of IBG changes the original power-angle relationship into a nonsinusoidal curve, introduces a limit on the peak of the rotor swing, and therefore brings about a new bottleneck for the transient stability.**

## I. Introduction

THE transient response of inverter-based generation (IBG) significantly differs from that of conventional power systems. When subjected to severe disturbances such as grid faults, the current-controlled mode of IBG should be implemented to avoid overcurrent damage and/or provide reactive current support for the power grid. The characteristics of the current-controlled mode are substantially different from the voltage-source characteristics dominated by synchronous generator (SG). The transient stability of conventional power systems has been extensively studied. Recent research efforts have also investigated the transient stability of grid-tied IBG systems [1], [2]. It was reported that the transient stability characteristics of these two types of systems are physically different [1].

With the growing penetration of IBG, the stability characteristics of modern power systems are jointly dominated by SGs and IBGs. Considering heterogeneous electrical sources including SGs and IBGs in the power system, the definition of "transient stability" can be extended to describe whether some of electrical sources lose synchronism from the others after a large disturbance. In the literature, few theoretical studies have been reported on the transient stability of the modern power system containing both SGs and IBGs. The transient stability mechanism remains to be elucidated, which entails analytical studies. Prior studies, however, were mostly based on simulations [3], [4], which can neither reveal the physical mechanism behind the transient instability phenomenon nor provide accurate information on the transient stable boundary.

The contribution of this letter is to analytically investigate the impact of IBG on the power system transient stability and provide a quantitative method to accurately predict the transient stability boundary. It is found that the sinusoidal power-angle relationship of the original system is altered by the integration of IBG. Moreover, the maximum of the SG's rotor-angle movement is limited by the synchronization of phase-locked loop (PLL). Therefore, the integration of IBG creates a new bottleneck hindering the system transient stability. The contribution provides analytical insights for reassessing the transient stability and relay protection of modern power systems.

## II. System Overview

Regardless of conventional power systems or grid-tied IBG systems, the fundamental mechanism of transient stability can be understood with a single-machine/converter infinite bus (SMIB/SCIB) system. For multi-machine/multi-converter power systems, the transient stability/instability phenomena can be qualitatively interpreted with the knowledge gained from the SMIB/SCIB system. To investigate the impact of IBG on the transient stability of the convention power system, an IBG is connected into the SMIB system, forming a simple prototype system, as shown in Fig. 1(a). The system is composed of a current-controlled IBG, a SG, and an infinite bus. Through a profound analysis towards such a simple system, the fundamental mechanism of transient instability can be clearly revealed, and a basic stability analysis approach can be developed. These basic cognitions are of significant importance to analyze and interpret the transient stability/instability phenomena in complex power systems.

Several basic assumptions are highlighted before modeling the system. i) The dynamics of current control loop (hundreds

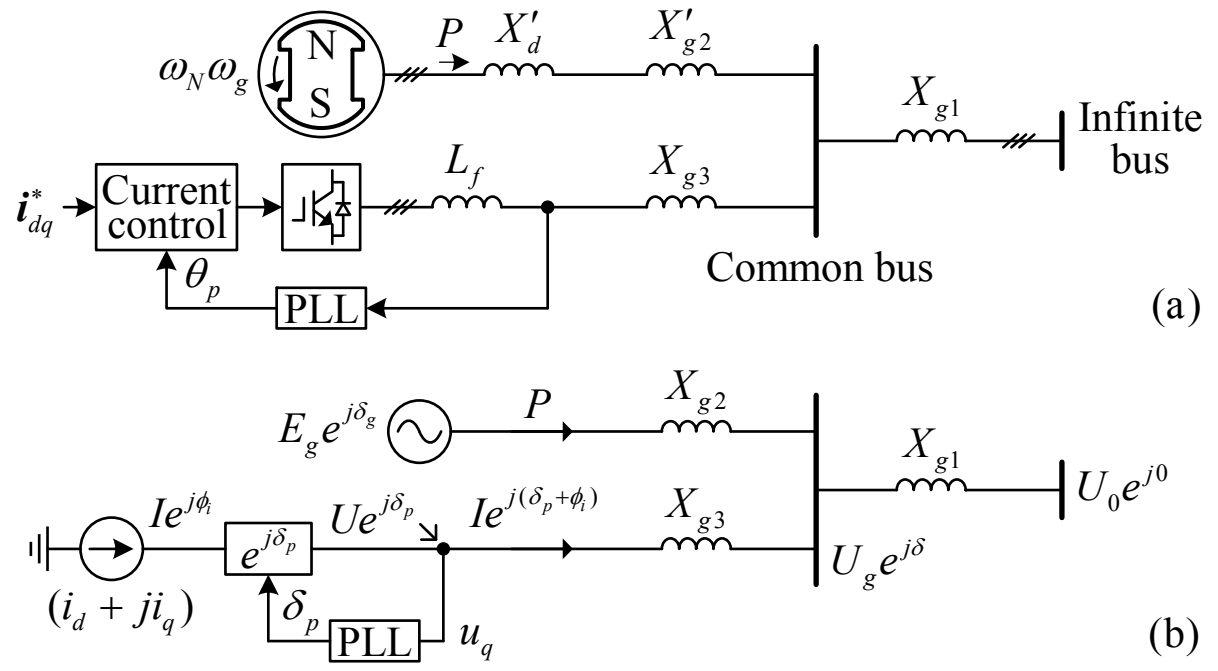

Fig. 1. (a) A current-controlled IBG is connected to the conventional SMIB system. (b) Simplified circuit. Note that the series connection of the current source with $X_{g3}$ is allowed since the current dynamics are neglected.

This work was supported by National Natural Science Foundation of China (NSFC) under Grant 61722307, Grant 52061635102. *(Corresponding author: Hua Geng.)*

The authors are with the Department of Automation, Beijing National Research Center for Information Science and Technology, Tsinghua University, Beijing, 100084, China (e-mail: he-xq16@mails.tsinghua.edu.cn; genghua@tsinghua.edu.cn).

of hertz) and PLL (tens of hertz) are overlooked, considering the timescale of interest in this study is focused on the rotor motion dynamics (typically 0.1 ~ 2 hertz). The PLL model is represented by algebraic equation [1]. ii) The electromagnetic transients in the system circuit [see Fig. 1(b)] are neglected. iii) The voltage-behind-reactance of the SG remains constant; iv) The resistive component of the system circuit is neglected. In view of the assumptions i) and ii), the current source representing the IBG can be connected in series with the inductance $X_{g3}$ [5]. The modeling in this study is based on the per-unit system.

## III. Modeling and Stability Analysis

### A. Stability Analysis of the PLL

Applying the superposition principle to Fig. 1(b), we can obtain the voltage equation at the synchronous reference frame,

$$U_g e^{j\delta} = K_1 E_g e^{j\delta_g} + K_2 U_0 e^{j0} + jX_{g4} I e^{j(\delta_p+\phi_i)} \quad (1)$$

where $K_1 = X_{g1}/(X_{g1}+X_{g2})$ , $K_2 = X_{g2}/(X_{g1}+X_{g2})$ , and $X_{g4} = X_{g1}X_{g2}/(X_{g1}+X_{g2})$. Fig. 1(b) explains other variables. Given $\delta_g \in [0, \pi]$, we can obtain that $\delta_p \in [0, 3\pi/2]$ according to $i_d = I\cos\phi_i \geq 0$. Note that it is assumed that the IBG outputs non-negative active current, regardless of normal or fault conditions [2], [5].

The terminal voltage equation of the IBG is

$$(u_d + ju_q)e^{j\delta_p} = K_1 E_g e^{j\delta_g} + K_2 U_0 e^{j0} + jX_{g5} I e^{j(\delta_p+\phi_i)} \quad (2)$$

where $X_{g5} = X_{g3} + X_{g4}$. Neglecting the PLL dynamics implies

$$u_q = K_1 E_g \sin(\delta_g - \delta_p) - K_2 U_0 \sin\delta_p + X_{g5} i_d = 0 \quad (3)$$

which determines the real-time relation between $\delta_p$ and $\delta_g$.

To achieve the grid-synchronization of the IBG, i.e., (3) must be solvable, the existence of the PLL's equilibrium point raises the following requirement,

$$\sqrt{(K_1E_g)^2 + (K_2U_0)^2 + 2K_1K_2E_gU_0\cos\delta_g} \geq X_{g5} i_d. \quad (4)$$

i. If $i_d \leq |K_1E_g - K_2U_0|/X_{g5}$ , (4) always holds, $\delta_{g\max} = \pi$ .
ii. If $|K_1E_g - K_2U_0|/X_{g5} < i_d \leq (K_1E_g + K_2U_0)/X_{g5}$ ,

$$\delta_{g\max} = \arccos\frac{(X_{g5}i_d)^2 - (K_1E_g)^2 - (K_2U_0)^2}{2K_1K_2E_gU_0} \in [0, \pi). \quad (5)$$

iii. If $i_d > (K_1E_g + K_2U_0)/X_{g5}$ , (4) [or (3)] never holds.

*Remark 1:* Only if $i_d \leq |K_1E_g - K_2U_0|/X_{g5}$ , $\delta_g \in [0, \delta_{g\max}]$ , then (4) can hold and (3) is solvable. In other words, the PLL synchronization, which is faster than the SG's rotor motion, limits the maximum rotor angle of the SG. When the rotor angle exceeds the maximum allowable power angle (MAPA), (3) becomes singular and the PLL loses synchronism due to the loss of equilibrium point.

Given a specific $\delta_g \in [0, \delta_{g\max}]$ , there are two solutions for $\delta_p$ to meet (3). The stable one can be identified by the negative feedback condition of the PLL $du_q/\delta_p < 0$ [5], i.e.,

$$K_1E_g\cos(\delta_g - \delta_p) + K_2U_0\cos\delta_p > 0. \quad (6)$$

### B. Power-Angle Relationship Analysis of the SG

The output power of the SG is

$$P + jQ = E_g e^{j\delta_g}[(E_g e^{j\delta_g} - U_g e^{j\delta})/(jX_{g2})]^*. \quad (7)$$

By recalling (1), we can obtain the active power component,

$$P(\delta_g) = P_0(\delta_g) - K_1E_gI\cos(\delta_g - \delta_p - \phi_i) \quad (8)$$

where $P_0(\delta_g) = E_gU_0\sin\delta_g/(X_{g1}+X_{g2})$ represents the SG's active power output when the IBG is disconnected. The IBG output current produces the last term in (8), which, therefore, alters the sinusoidal power-angle relationship $P_0(\delta_g)$ .

*1) Pre-Fault and Post-Fault Condition:* It can be considered that the IBG purely outputs active current in normal conditions. $\phi_i = 0$ , $P(\delta_g) = P_0(\delta_g) - K_1E_gI\cos(\delta_g - \delta_p)$ . Proposition 1 below indicates the difference between $P(\delta_g)$ and $P_0(\delta_g)$ .

*Proposition 1:* If and only if $K_1E_g + X_{g5}i_d < K_2U_0$ , $P(\delta_g) > P_0(\delta_g)$ can hold at a point $\delta_g \in [0, \pi]$ .

*Proof:* Zeroing the last term in (8) and considering (3) and (6) lead to $\cos\delta_g = -(K_1E_g + X_{g5}i_d)/(K_2U_0)$ . There exists a solution $\delta_{\text{gcross}}$ within $(0, \pi)$ if and only if $K_1E_g + X_{g5}i_d < K_2U_0$ . We can further verify $P(0) < P_0(0)$ . Hence, the proof is completed by the continuity of the functions $P(\delta_g)$ and $P_0(\delta_g)$ .□

*Remark 2:* Generally, the power-angle curve of the SG is lowered in normal conditions due to the impact of the IBG output current. This is because the IBG's active current output drops the common bus voltage, as denoted by the last term in (1). Only if $K_1E_g + X_{g5}i_d < K_2U_0$ , the power-angle curve segment within the range $(\delta_{\text{gcross}}, \pi]$ is elevated.

*2) Fault-On Condition:* If the IBG is requested to purely outputs reactive current to support the common bus voltage during grid faults, it can be obtained that $\phi_i = -\pi/2$ [5], $P(\delta_g) = P_0(\delta_g) + K_2U_0I\sin\delta_p$ , where $0 \leq \delta_p \leq \delta_g \leq \pi$ considering (3) and (6). The last term in $P(\delta_g)$ is greater than zero except at 0 and $\pi$ points. This suggests that the reactive current support from the IBG can improve the active power output of the SG during the fault-on period.

### C. Transient Stability Analysis of the Entire System

The rotor swing equation of the SG is

$$\begin{aligned} d\delta_g/dt &= \omega_N(\omega_g - \omega_0) \\ T_J\, d\omega_g/dt &= P_m - P - D(\omega_g - \omega_0). \end{aligned} \quad (9)$$

For the conventional SMIB system, its transient stability boundary is dominated by the unstable equilibrium point (UEP). The stable manifold of the UEP forms the stability boundary [6]. For the SMIB system connected with an IBG, considering the fact that there may be no UEP within $[0, \delta_{g\max}]$, a separate discussion is necessary.

*1) With UEP:* In this case, as shown in Fig. 2(a), the stability boundary is still dominated by the UEP. The exact stability region can be found using the method developed in [6], which is based on the backward integral from the UEP.

*2) Without UEP:* In this case, the stability boundary is taken over by the MAPA $\delta_{g\max}$ , as shown in Fig. 2(b) or (c). After the fault clearance, the critical stable scenario is that, when the power angle increases to $\delta_{g\max}$ on the first swing, the rotor speed decreases to zero. To predict the stability boundary in this case, we can numerically integrate (9) backward from the critical state $(\delta_g, \omega_g) = (\delta_{g\max}, 0)$ . The resulting state trajectory is the transient stability boundary.

Fig. 2 plots the power-angle curve and stability boundary in the following three cases:

i. Case A: $I$ = 0.5 pu, $P_m$ = 1.0 pu, there is a UEP.

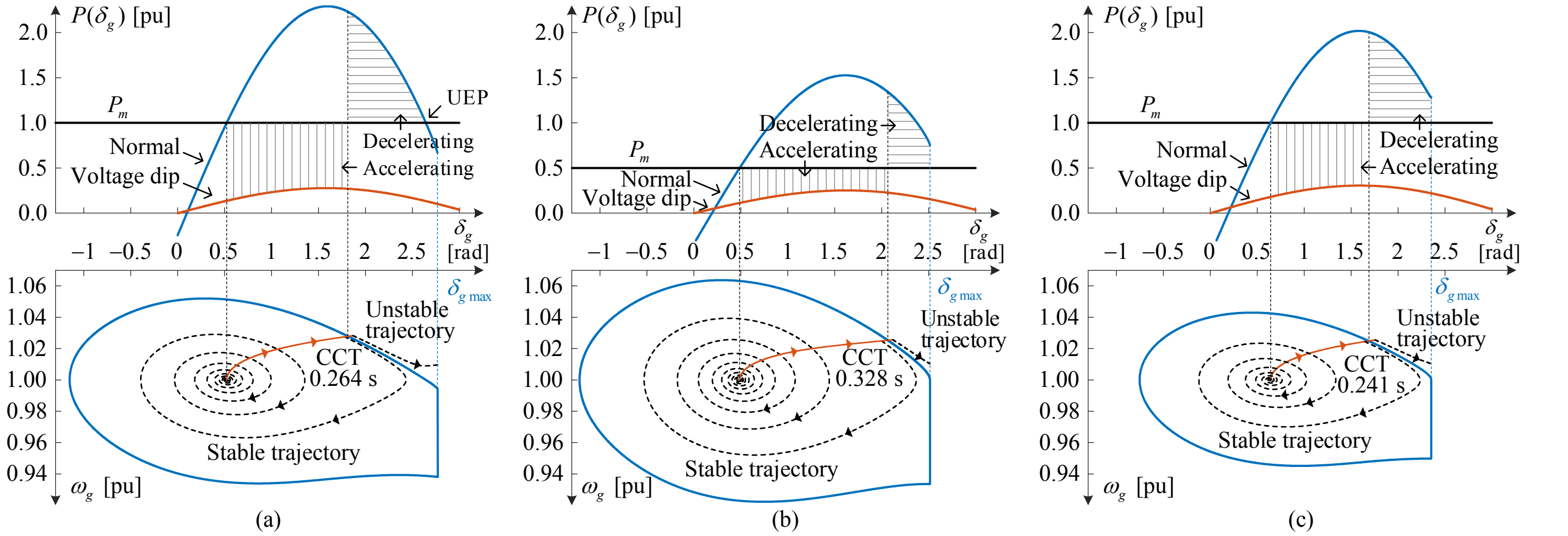

Fig. 2. Power-angle curve, stability boundary, and stable and unstable trajectories in three cases. (a) Case A: $I = 0.5$ pu, $P_m = 1.0$ pu. (b) Case B: $I = 1.0$ pu, $P_m = 0.5$ pu. (c) Case C: $I = 1.0$ pu, $P_m = 1.0$ pu. There is a UEP in Case A whereas there is no UEP in Cases B and C.

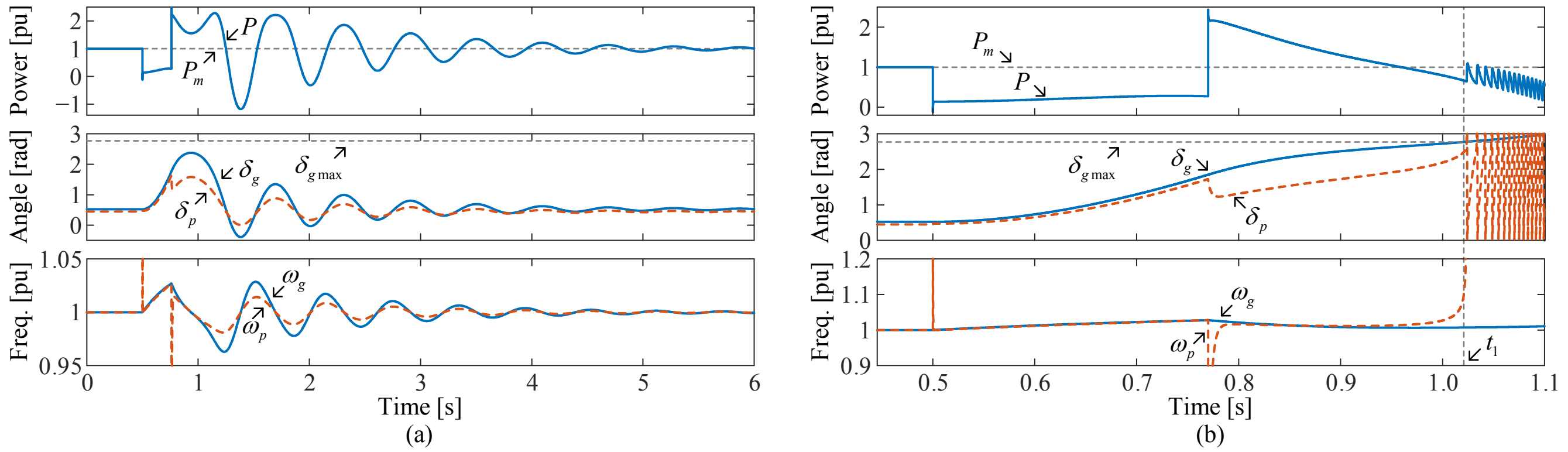

Fig. 3. The grid voltage dips at 0.5 s. The system is stable or unstable when the voltage dip lasts 0.26 s in (a) or 0.27 s in (b), respectively. In (b), when the power angle $\delta_g$ surpasses $\delta_{gmax}$ at $t_1$, the system collapses since the PLL loses synchronism rapidly. After that the SG goes out-of-step gradually.

ii. Case B: $I = 1.0$ pu, $P_m = 0.5$ pu, there is no UEP.
iii. Case C: $I = 1.0$ pu, $P_m = 1.0$ pu, there is no UEP.

The system parameters are summarized in Appendix. It is shown in Fig. 2 that after the grid voltage dips, the power imbalance drives the SG's rotor to accelerate. If the voltage-dip fault is cleared before the critical clearing time (CCT), the rotor will undergo a swing process to converge. If the fault is cleared after the CCT, the rotor will diverge. Once the MAPA is reached, the transient instability occurs. The stable and unstable trajectories are shown in Fig. 2, where the correctness of the stability boundaries is verified.

The CCT has also been verified by simulations, where the IBG model includes the dynamics of the PLL and current control. Considering the CCT = 0.264 s in Fig. 2(a) as an example, the verification results are shown in Fig. 3. Fig. 3(b) displays that once the power angle crosses over the MAPA $\delta_{g\max}$, the system collapses since the PLL loses synchronism.

*Remark 3:* When the grid-connected IBG's capacity is small, there is a UEP and it is the bottleneck that determines the transient stability boundary. The transient instability mechanism in this case remains unchanged compared to the SMIB system. When the IBG's capacity is high enough, the UEP disappears. The MAPA on the power-angle curve decides the maximum movement of rotor angle and consequently becomes the new bottleneck impeding the transient stability. In this case, the transient instability mechanism is linked to the PLL synchronization introduced limit on the power angle. This new instability mechanism is intrinsically different from the instability associated with the UEP. It is also suggested that the MAPA leads to a new factor affecting the CCT and therefore system operators need to pay attention to the potential impact of the IBG's PLL on the relay protection of power systems.

It should be noted that even if the IBG is disconnected from the grid when the PLL loses synchronism, the synchronization of the SG cannot be guaranteed because the UEP will reappear and bring about a new stability boundary for the rest of the system. However, the decelerating area will increase due to the disconnection of the IBG, which will be beneficial for the SG's synchronization.

This study investigates the impact of the current-controlled IBG on the transient stability of the SG, in which the effect of the dynamics of PLL and current control loop is overlooked. It should be mentioned that additional potential stability issues may be caused by PLL and current control loop, which have been reported in [5] and [7].

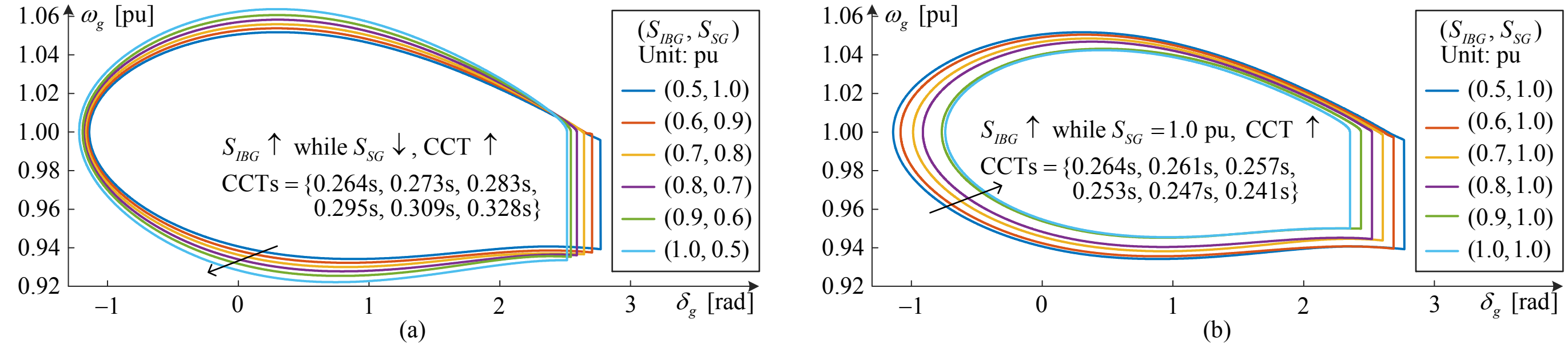

Fig. 4. (a) With the increase of the proportion of the IBG's capacity (the total system capacity remains 1.5 pu), the CCT increases while the maximum allowable power angle (MAPA) decreases. (b) With the increase of the IBG's capacity (the SG's capacity remains 1.0 pu), both the CCT and the MAPA decreases. Note that $S_{IBG}$ and $S_{SG}$ in the figure represent the IBG's and SG's capacity, respectively.

### *D. Impact of Capacity Proportion on the Transient Stability*

In Fig. 2, it is implied that the capacity change of the SG and IBG has different impact on the transient stability margin. Overall, the order of transient stability margin of the three cases from low to high is Case C < Case A < Case B.

The comparison between Cases A and B indicates that under the condition of maintaining the total capacity constant, the higher the IBG's capacity proportion is, the larger the CCT is. By comparison with the case of a large SG's capacity in Fig. 2(a), it is displayed in Fig. 2(b) that a small SG's capacity leads to a small accelerating and decelerating area. More importantly, the accelerating region is elongated, which increases the critical clearance angle (CCA) and accordingly the CCT, although the inertia is reduced with the SG's capacity. This observation is also verified by the result in Fig. 4(a), where the changes in the stability boundary and CCT with the IBG's capacity proportion are displayed. Moreover, it is found that the MAPA is reduced with the increment of the IBG's capacity.

By comparison between Cases A and C, it is revealed that while maintaining the SG's capacity constant, the increase of the IBG's capacity impairs the transient stability margin. This is essentially because a high IBG's capacity leads the power-angle curve to shift downward and also reduces the MAPA. This observation is also verified by the result in Fig. 4(b).

*Remark 4:* One must be cautious to interpret the results in this subsection. On one hand, the validity of the results relies on the assumptions in Section II; the results may also vary with different system parameters and operating points. On the other hand, the transient stability of multi-machine/multi-converter power systems is rather complicated. The generalization of the results still needs a profound analysis in the future.

## IV. Conclusion

Regarding the impact of IBG on the transient stability of the SMIB system, the analytical insights into the impact mechanism are provided in this letter, and a method to quantify the accurate stability region is also developed. With the integration of IBG, the sinusoidal power-angle curve of the SMIB system is altered and the maximum allowable power angle of the SG is constrained. The power angle constraint creates a new bottleneck hindering the transient stability. The transient stability margin and CCT are significantly affected by the proportion of the IBG's capacity. The modeling and analysis in this letter are also applicable to power-consumption converter equipment such as the rectifier of VSC-HVDC. Based on the findings in this letter, future efforts will be devoted to the research on the transient stability of complex power systems.

## Appendix

| |
|---|
| The base of the per-unit system: |
| $S_N$ = 10 MVA, $U_N$ = 690 V, $\omega_N = 100\pi$ rad/s; |
| $X_{g1}$ = 0.2 pu, $X_{g2}'$ = 0.1 pu, $X_{g3}$ = 0.3 pu; |
| $E_g$ = 1.15 pu, $U_0$ = 1.0 or 0.1 pu; $\phi_i = 0$ or $-\pi/2$; |
| $X_d'$, $T_J$, and $D$ are subject to the SG's capacity. For a 1.0 pu capacity (i.e., 10 MVA), $X_d'$ = 0.16 pu, $T_J$ = 6 s, $D$ = 10. |

## References

[1] W. Tang, J. Hu, Y. Chang, X. Yuan, and F. Liu, "Modeling of DFIG-based WT for system transient response analysis in rotor speed control timescale," *IEEE Trans. Power Syst.*, vol. 33, no. 6, pp. 6795–6805, Nov. 2018.

[2] X. He, H. Geng, R. Li, and B. C. Pal, "Transient stability analysis and enhancement of renewable energy conversion system during LVRT," *IEEE Trans. Sustain. Energy*, vol. 11, no. 3, pp. 1612–1623, Jul. 2020.

[3] H. N. Villegas Pico and B. B. Johnson, "Transient stability assessment of multi-machine multi-converter power systems," *IEEE Trans. Power Syst.*, vol. 34, no. 5, pp. 3504–3514, Sep. 2019.

[4] A. A. van der Meer, M. Ndreko, M. Gibescu and M. A. M. M. van der Meijden, "The effect of FRT behavior of VSC-HVdc-connected offshore wind power plants on AC/DC system dynamics," *IEEE Trans. Power Del.*, vol. 31, no. 2, pp. 878–887, Apr. 2016.

[5] X. He, H. Geng, J. Xi, and J. M. Guerrero, "Resynchronization analysis and improvement of grid-connected VSCs during grid faults," *IEEE J. Emerg. Sel. Top. Power Electron.*, in press.

[6] H. D. Chiang, M. Hirsch, and F. Wu, "Stability regions of nonlinear autonomous dynamical systems," *IEEE Trans. Automatic Control*, vol. 33, no. 1, pp. 16–27, Jan. 1988.

[7] J. Chen, M. Liu, T. O'Donnell, F. Milano, "Impact of current transients on the synchronization stability assessment of grid-feeding converters," *IEEE Trans. Power Syst.*, vol. 35, no. 5, pp. 4131–4134, Sep. 2020.